\documentclass[modern]{aastex63} 

\lefthead{Wada, Tsukamoto, Kokubo}
\righthead{Planets around SMBH}

\newcommand{\bea}{\begin{eqnarray} }
\newcommand{\eea}{\end{eqnarray}}

\begin{document}


\title{Planet Formation around Super Massive Black Holes in the Active Galactic Nuclei}
 
\author{%
Keiichi Wada
}
\affiliation{Kagoshima University, Graduate School of Science and Engineering, Kagoshima 890-0065, Japan}
\affiliation{Ehime University, Research Center for Space and Cosmic Evolution, Matsuyama 790-8577, Japan}
\affiliation{Hokkaido University, Faculty of Science, Sapporo 060-0810, Japan}
\correspondingauthor{Keiichi Wada}
\email{wada@astrophysics.jp}

\author{
Yusuke Tsukamoto
}%
\affiliation{Kagoshima University, Graduate School of Science and Engineering, Kagoshima 890-0065, Japan}

\author{Eiichiro Kokubo}%
\affiliation{National Astronomical Observatory of Japan, Mitaka 181-8588, Japan}


%


\begin{abstract}
As a natural consequence of the elementary processes of dust growth, 
we discovered that a new class of planets can be formed around supermassive black holes (SMBHs).
We investigated a growth path from sub-micron sized icy dust
 monomers to Earth-sized bodies outside  the ``snow line'', located several 
 parsecs from SMBHs in low luminosity active galactic nuclei (AGNs).
In contrast to protoplanetary disks, the ``radial drift barrier'' does not prevent the formation of
planetesimals.  In the early phase of the evolution, low collision velocity between dust particles promotes
 sticking; therefore, the internal density of the dust aggregates
 decreases with growth.
When the porous aggregate's size reaches 0.1--1 cm, the collisional compression becomes effective, and the decrease in
 internal density stops.
Once 10--100 m sized aggregates are formed, they are decoupled from gas 
 turbulence, and the aggregate layer becomes gravitationally unstable,
 leading to the formation of planets by the fragmentation of the layer, with ten times the mass of the earth.
The growth time scale depends on the turbulent  strength of the circumnuclear disk
and the black hole mass $M_{BH}$, and it is comparable to the AGN's lifetime ($\sim 10^8$ yr) for low mass ($M_{BH} \sim 10^6 M_\odot$) SMBHs.
\end{abstract}


\section{INTRODUCTION}
Planetary systems are ubiquitous -- more than four thousand exoplanets
 have been discovered thus far\footnote{https://exoplanetarchive.ipac.caltech.edu/}.
However, protoplanetary disks around stars may not be the only site for
 planet formation in the universe.
Here we propose a new site of ``planet" formation: the circunumnuclear disk around 
supermassive black holes (SMBHs). 
 
 Most galaxies host SMBHs at their centers, with masses ranging from a
 few million to billion solar masses.
Gas disks around SMBHs emit large amount of energy owing to mass accretion
 onto the SMBHs, which are known as the ``central engine'' of active
 galactic nuclei (AGNs). 
It is believed that the mass of SMBH in a galaxy depends on its host galaxy's bulge mass \citep{marconi2003}. 
Researchers are more convinced of the presence of SMBHs since 
the discovery of the ``black hole shadow" in  M87 \citep{eht2019}.
In the ``unified model" of AGNs \citep{antonucci1993, netzer2015},
the gas and dust form a geometrically and optically thick ``torus", and it obscures the broad
  emission line (line width is several 1000 km s$^{-1}$) region around the central accretion disk. 
  This hypothesis successfully explains the type-1 and type-2 dichotomy of Seyfert galaxies' spectra,
depending on the viewing angle of the tori.
The real structure of the tori has been unclear for many years. Recently, the  Atacama Large Millimeter/sub millimeter Array (ALMA) spatially resolved
the molecular tori in nearby AGNs \citep{garcia-burillo2016, imanishi2018, izumi2018, combes2019}.
Their internal structure is still not well resolved; 
however, recent 3-D radiation-hydrodynamic simulations suggested 
a dynamic structure energized by a radiation-driven fountain flow to sustain their geometrical thickness \citep[][see also Fig. 1]{wada2012, wada2018}.
Notably, even in this situation, cold, dense gas forms a geometrically thin disk \citep{schartmann2014, wada2016},
and this stratified structure is also consistent with recent X-ray surveys \citep{buchner2014}.

 The remainder of this paper is organized as follows. In \S 2, we describe dust and its environment around SMBHs, 
 and their differences from the standard situation, i.e.  in the circum-stellar disks.
  In \S 3, we show a typical evolutional track of a representative 
 dust particles from a monomer to a planet-sized body. 
 Four stages of the dust coagulation based on the recent theoretical model proposed for
 the proto-planet disks are described in details in Appendix.
  We also discuss how the evolutional time scales
 depend on parameters in \S 5.


\begin{figure}[h]
\begin{center}
\includegraphics[width = 10cm]{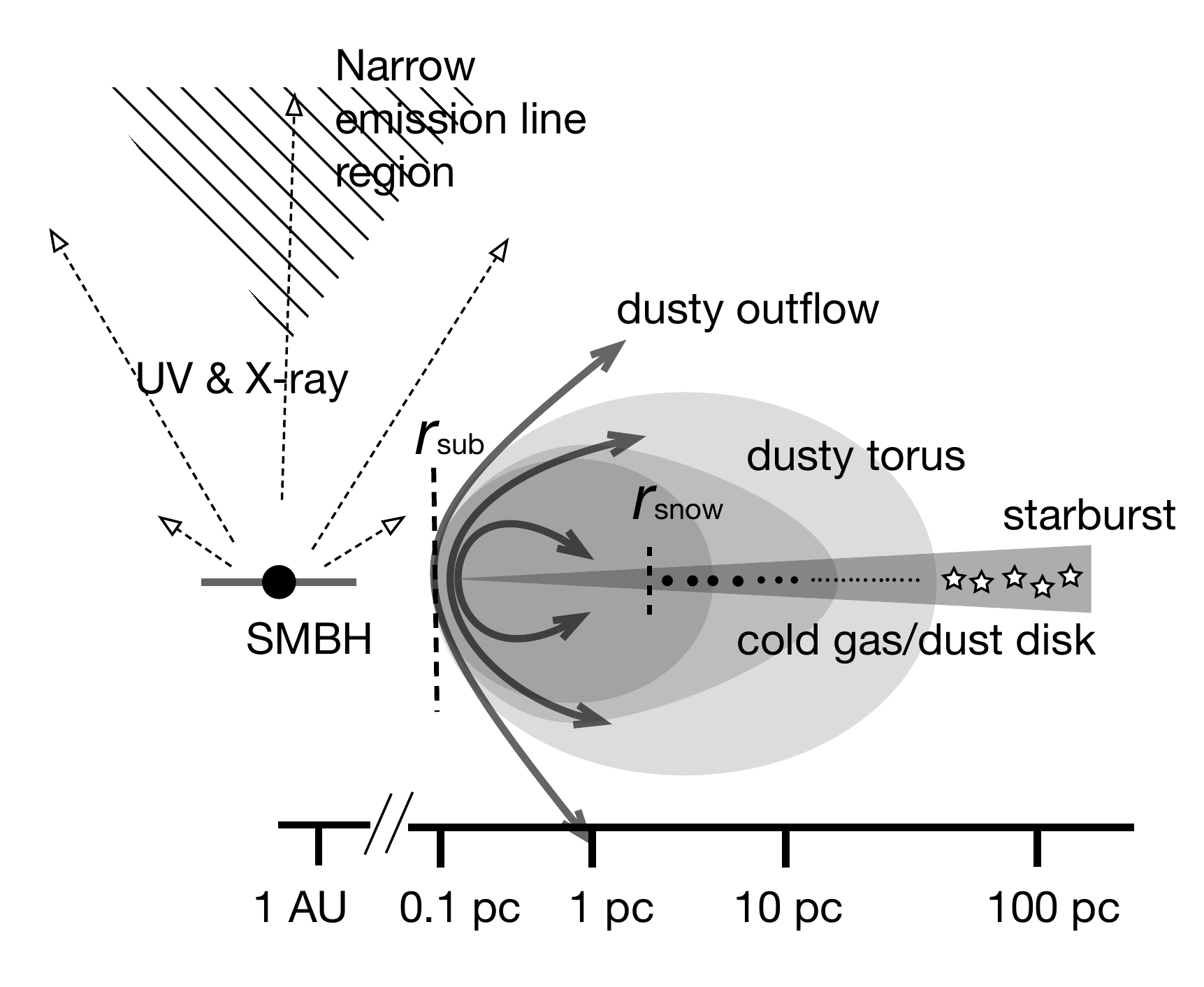}    
\caption{A schematic picture of the Active Galactic Nucleus (AGN) and the circumnuclear disk. A supermassive black hole (the mass is $10^6-10^9 M_\odot$) is surrounded by an accretion disk, which radiates enormous energy ($\sim 10^{42}- 10^{45}$ erg s$^{-1}$) mostly in the ultra-violet and X-ray.  The dust particles in the central $r < r_{sub} \sim $ 0.1- a few parsecs are sublimated owing to the heating by the central radiation.
The radiation forms conical ionized gas (Narrow emission-line region) and also contributes to producing outflows of the dusty gas and torus \citep{wada2012, wada2018, izumi2018}.
In the mid-plane of the torus, cold, dense gas forms a thin disk, where icy dust particles can present beyond the snow-line $r_{snow}$. The dust aggregates evolve by collisions to form planetesimals, and
eventually ``circum-black hole planets" by the gravitational instability of the aggregate disk. 
}
\end{center}
\label{fig:1}
\end{figure}

\begin{table}
\centering
\caption{Differences between the proto-planetary disk and AGN.  }
\medskip
\begin{tabular}{l | l |l}
\hline
 & proto-planetary disk & circumnuclear disk  \\
\hline
mass of the central object & $M_\star \sim M_\odot$ &   $M_{BH} \sim 10^{6-9} M_\odot$ \\
lumninosity of the central source & $\sim L_\odot$& $10^{10-12} L_\odot$ \\
spectrum of the central source & black body  &  power law \\
size of the dusty disk & 10-100 AU & 0.1 pc - 100 pc \\
inner edge of the dusty  disk&  $\sim$ 0.1 pc$^a$  &   dust sublimation radius (sub pc $\sim$ pc) \\
gas mass & $\sim $ 0.01 M$_\star$ & $\sim$ 0.1 $M_{BH}$ \\
dust mass & $\sim 10^{-4} M_\odot $  &  $\sim 10^{3} - 10^6 M_\odot $ \\
rotational period & $\sim$100 yr & $\sim 10^6$- $10^8$ yr  \\
life time & $\sim 10^6$ yr & $10^7 - 10^8$ yr \\
drag law &   Epstein/Stokes      &  Epstein  \\
mean free path of gas&   $\sim $1-100 cm    & $\sim 10^{12}(\frac{n}{10^3 \, {\rm cm} \, {\rm s}^{-3}} )^{-1}$ cm \\
\hline

\end{tabular}
$a$ \citet{eisner2007, suzuki2010} 
\end{table}

%
\section{Environment of dust around AGNs in comparison with proto-planetary disks}

\subsection{The snow line in AGNs and major difference from the proto-planetary disks}
The dusty gas around SMBHs extends beyond the dust sublimation radius $r_{sub}$, where the dust temperature is
higher than the sublimation temperature of the dust grains ($T_{sub} \sim$ 1500 K). 
The radius  depends on the AGN luminosity: $r_{sub} =1.3 \, {\rm pc} \left( \frac{L_{UV}}{10^{43} \, {\rm erg}\, s^{-1}}\,  \right)^{0.5} \; \left(\frac{T_{sub}}{1500 \, K} \right)^{-2.8}  
 \; \left(\frac{a_{d}}{1 \, \mu m} \right)^{-1/2}$, where $L_{UV}$ is the ultra-violet luminosity of the AGN,  and $a_d$ is the dust size  \citep{barvainis1987}.
 The temperature of the gas and dust beyond $r_{sub}$, especially at the
 mid-plane of the dusty torus, should be cold $\lesssim 100$ K \citep{schartmann2014}, because the radiation originated from the accretion disk
 is weaker in the direction of the disk plane, and
it is further attenuated by the dense dusty gas. 
Interestingly, even for X-rays, a large fraction of AGNs are Compton-thick, i.e.
the Hydrogen column density is  $N_H >10^{24}$ cm$^{-2}$ \citep{buchner2014}. 
Although near-infrared and mid-infrared interferometer observations of AGNs show the presence of hot dusts (several 100 K) around AGNs \citep{tristram2014},
colder dust particles are also present in this dense media around AGNs. 
The total amount of dust in the central 6-27 pc around SMBHs estimated 
from recent  molecular lines (e.g. CO) observations of nearby AGNs by ALMA \citep{combes2019}
is enormous, e.g. $\sim 0.7-3.9 \times 10^5 M_\odot$ for the dust-to-gas mass ratio of $\sim 0.01$ \citep{draine2011}. 
This number could be even larger for the high metallicity environment around AGNs \citep{groves2006, remy2014}.
The internal dynamics and structure of the molecular tori are still observationally unclear.
However, since the mass feeding to the AGN through the circumnuclear disk 
is necessary during their lifetime ($\sim 10^7-10^8$ yr), 
the turbulent viscosity works in the dusty gas disk.
Here, we model the turbulent disk based on the $\alpha$-viscosity formalism \citep{shakura1973, shlosman1987}.

The dust grains in the circumnuclear disks around SMBHs are in a qualitatively similar situation as the ones in 
 proto-planetary disk.  The major difference between the two systems are summarized in Table 1.
The ``snow line" for $a_d = 0.1 \mu m$ dust irradiated by X-ray around an AGN with a SMBH of $10^7 M_\odot$ for the Eddington ratio ($\gamma_{Edd}$) of 0.1 is
  \begin{eqnarray}
  r_{snow} \approx  4.7  \, {\rm pc}  (L_X/1.3\times 10^{42} \, {\rm erg \,s}^{-1}) ^{1/2}  (T_{ice}/170 \, {\rm K})^{-2.8} \, a_{d, 0.1}^{-1/2}
  \end{eqnarray}  
  
  \citep{barvainis1987} \footnote{The approximate proportionality $a_d^{-1/2}$ comes from the
absorption efficiency of a dust grain being roughly proportional to its radius at a certain wavelength in the near-IR \citep{draine1984}.}.
  Moreover, AGNs are often heavily obscured (Compton-thick) even for hard-X rays \citep{buchner2014}, suggesting that
 a cold dusty layer exits around SMBHs. 
   Therefore, it  is expected that the dust in the most part of the circumnuclear disk is icy.  

\subsection{Outline of evolution of ``fluffy" dust aggregates}
  
  We then apply recent models of coagulation of dust particles and their aggregates outside the snow line 
  in the protoplanetary disk \citep{okuzumi2012, kataoka2013, suyama2012, michikoshi2016, michikoshi2017} to the dust around AGNs.
  The coagulation of 'fluffy icy dust' is one of the plausible solutions to avoid the theoretical obstacles that prevent from growth of dust grains (monomers) to ``planetesimal" such as the ``radial drift barrier" \citep{okuzumi2012}.
  In this scenario, the evolution of the dust can be divided into four stages:
  1)  ``hit-and-stick" phase, 2) collisional or gas pressure compression phase, 3) gravitationally compression phase, 
  and 4) gravitational instability phase \citep[e.g.][]{goldreich1973}. We investigated each phase in the circumnuclear disk as discussed below (see also Appendix).
  
 We track the evolution of icy monomers, whose size and density are
$a_0 = 0.1 \, \mu m$ and $\rho_0 = 1$ g cm$^{-1}$, and their aggregates.
 Here, we investigate the evolution of a representative dust particle size, using the single-size approximation \citep{sato2016}.
 In the hit-and-stick phase, when two monomers/aggregates collide, 
 the internal structure of the aggregates becomes porous (i.e. average internal density is smaller than $\rho_0$)
 with internal voids \citep{suyama2012}.  This``fluffy dust" formation is also examined by numerical
 experiments \citep{dominik1997, wada-k2008a}.  The internal density and size of the aggregates are
$ \rho_{int} \sim (m_d/m_0)^{-1/2} \rho_0$ and $a_d \sim (m_d/m_0)^{1/2} a_0$, where
$m_d$ and $m_0$ are the masses of the aggregate and monomers, for 
the fractal dimension of 2.  When the collision energy exceeds a critical value,
the porous aggregates start to get compressed, and the evolution of the internal density changes beyond this point \citep{suyama2012}.
 During this compression phase, 
the aggregates' mass rapidly increase; however, their
internal densities gradually increase as well, from $\rho_{int} \sim 10^{-6}$ g cm$^{-3}$ to $\sim 10^{-4}$ g cm$^{-3}$.

In contrast to the dust coagulation process in the protoplanetary disks \citep{weidenschilling1977}, 
the drag between dust particles and gas obeys the Epstein law  only.  
The aggregate's size ($a_{d}$) is always much smaller than the mean free path of the gas ($\lambda_{mfp}
\sim   10^{12} \; {\rm cm} \left( \frac{\sigma_{mol}}{10^{-15} \, {\rm cm}^2} \right)^{-1}
\left( \frac{n_{mol}}{ 10^3 \, {\rm cm}^{-3}} \right)^{-1}$, where $\sigma_{mol}$ and $n_{mol}$ are the collisional cross-section and
number density of the gas).
At all times the radial drift velocity of the dust is
negligibly small compared to the Kepler velocity $v_K$ (i.e. $10^{-4}- 10^{-5} v_K$).

In both the protostellar and the circumnuclear disks, the dust-gas coupling is characterized by the normalized stopping time, 
i.e., the Stokes number,  $S_t \equiv \Omega_K \, t_{stop}$, where $t_{stop}$ is  the time scale of the dust particles to reach the terminal velocity due to the gas drag.
In the Epstein law, $t_{stop}$ is proportional to $\rho_{int} \,a_d$, then $S_t$ is 
\begin{eqnarray}
S_t &=& \frac{\pi \rho_{int} \, a_d }{2 \Sigma_g}  \nonumber \\
&=& \frac{\pi \rho_{int} \, a_d (\pi G Q_g)}{ 2 c_s \Omega_K } \nonumber \\
&\sim&  1.5\times 10^{-5}  \, \rho_{int,1}  \, a_{d,0.1} \, c_{s,1}^{-1} \, r_1^{3/2} \, M_{BH,6}^{-1/2} \,Q_g,
\label{eq:stokes}
\end{eqnarray}
 where $Q_g$ is the Toomre's Q-value for a gas disk and $Q_g \equiv  c_s \Omega_K/(\pi G \Sigma_g)$, with the surface density
 of the gas disk $\Sigma_g$, and $a_{d,0.1} \equiv a_d/(0.1\,  \mu m)$, the sound velocity of the gas $c_{s,1} = c_s/ (1 \, {\rm km} \, {\rm s}^{-1}$)  and $\rho_{int, 1} \equiv  \rho_{int}/(1 \, {\rm g} \, {\rm cm}^{-3})$.
$Q_g < 1$ is a necessary condition for the ring-mode gravitational instability.

The radial velocity of the dust $v_{r, d}$ relative to the gas \citep{weidenschilling1977, tsukamoto2017} is
\begin{eqnarray}
v_{r, d} =  \frac{2 S_t}{1 + S_t^2} \, \eta \, v_K,
\label{eq-18}
\end{eqnarray}
where $\eta$ is a parameter that determines  the sub-Keplerian motion of the gas, 
\begin{eqnarray}
\eta \equiv  -\frac{1}{2} \frac{c_s^2}{ v_K^2} \frac{d \ln P}{d \ln r} \sim 2 \times 10^{-5} M_{BH, 7}^{-1} \, c_{s,1}^2,
\label{eq-19}
\end{eqnarray}
where $M_{BH,7} \equiv M_{BH}/10^7 M_\odot$. 
Therefore,  both in the early stage of  the dust evolution ($S_t \ll 1$) and in the late phase ($S_t \sim 1$)  in the circumnuclear disk, 
$v_{r, d}$ is much smaller than $v_K$, then we can ignore the radial drift of the aggregates in the circumnuclear disk during the whole evolution.
The ``radial drift barrier", i.e. the dust growth is limited by infalling to the
central stars before dust particles obtain large enough mass as planetesimals, 
 is not a serious problem in the circumnuclear disk.
 
 \subsection{The Growth time and destruction by collisions}

The  growth time of the aggregate during the hit-and-stick phase can be estimated as in   \citet{tsukamoto2017}:
\begin{eqnarray}
t_{grow} &\equiv &(d \ln m_d/dt)^{-1}  \label{eq: 6} \nonumber \\
 &=&  \frac{4\sqrt{2\pi}}{3}  \frac{H_d \, \rho_{int} \, a_d}{ \Delta v \, \Sigma_d}  =  \frac{8(2\pi)^{3/2}}{3}   \frac{H_g}{\sqrt{\alpha} \, R_e^{1/4} \, c_s \, f_{dg}}  \label{eq: 7} \nonumber \\
 &\sim&  6.3 \times 10^7 \; [{\rm yr}]  \;   c_{s,1}^{-1} \left( \frac{f_{dg}}{0.01} \right)^{-1} \left( \frac{H_g}{0.1 \, {\rm pc}}  \right)  \left( \frac{\alpha}{0.1}  \right)^{-1/2} \,   \left( \frac{R_e}{10^4}  \right)^{-1/4}  \label{eq: 5} 
\end{eqnarray} 
where, $f_{dg}$ is the dust-to-gas ratio and $H_d$ and $H_g$ are scale heights of dust and gas disks, and
$H_d \approx H_g \propto M_{BH}^{-1/2} r^{3/2}$  in the circumnulear disk.  The Reynolds number $R_e$ and the relative velocity of the 
dust particles $\Delta v$ are given in Appendix.
This growth time is comparable to the AGN life time (see \S 3 and Appendix in more details).

During the dust compression phase due to collisions, the kinematics of the dust aggregates are dominated by eddies of turbulence in the
gas disk. Therefore, relative velocity of the aggregates, $\Delta v$, which is important for both growth and destruction of them, is 
determined by the property of the turbulence, i.e., $R_e$, and $S_t$ \citep{ormel2007}.
The dimensionless parameter $\alpha$ is a parameter 
to determine the kinematic viscosity in the turbulent disk \citep{shakura1973}:
\begin{eqnarray}
\alpha &\equiv& \frac{\dot{M_g}}{3\pi \Sigma_g c_s^2/\Omega_K} = \frac{\dot{M_g} G}{3 c_s^3} Q_g \simeq 0.3 \, Q_g \left(\frac{\gamma_{Edd, 6}}{0.01} \right) c_{s, 1}^{-3},
\label{eq: alpha}
\end{eqnarray}
where $\dot{M_g}$ is the radial mass accretion rate in the disk, and $\gamma_{Edd, 6}$ is the Eddington rate for the BH mass with $M_{BH} = 10^6 M_\odot$
for the energy conversion efficiency of 0.1.
Here, we assume $\alpha $ is a constant, smaller than unity throughout the disk. 
In this phase,  $S_t$ gradually increases, and eventually the phase ends when $S_t \sim 1$, then the aggregates decouple from the
turbulent gas. At this time,  the mass and size of the aggregates become $m_d \sim 10^5$ g and $a_d \sim 100$ m, respectively.
There internal density is still very low (i.e., ``fluffy"). 

In the next gravitationally compression phase, 
the aggregates are compressed owing to their self-gravity, and their internal
density increases as $\rho_{int} \propto m_d^{0.4}$ \citep{kataoka2013}.
The relative velocity of the aggregates in this phase is determined by the 
energy balance among gravitational scattering, collisional energy loss, turbulent stirring, turbulent 
scattering and gas drag \citep{michikoshi2016, michikoshi2017} (see also Appendix).  
The aggregates finally grow to $\sim $ km-sized bodies (i.e., planetecimals). 

The value of the critical velocity for 
collisional destruction of cm to km sized dust aggregates is not clear.
Numerical experiments of collisions between aggregates \citep{wada-k2009} showed that  
the critical velocity is 50-100 ${\rm m} \, {\rm s}^{-1}$ 
for  $\sim 10^4$ monomers ($m_d \sim 10^{-11}$ g), 
and this scales with the mass of the aggregates as $\propto m_d^{1/4}$.
If the critical velocity simply scales, 
it should exceed  1 ${\rm km} \, {\rm s}^{-1}$ for
km-sized `planetesimals'. 

In the regime with $S_t > 1$, if the Toomore $Q$-value for the disk of aggregates becomes smaller than $\sim 2$, 
the gravitational instability takes place \citep[e.g.][]{goldreich1973}, and spiral density enhancements are formed, and
it leads to rapid growth of 
planetesimals in a rotational period  $t_K  \equiv 2\pi/\Omega_K \simeq 9.5 \times 10^4 \, {\rm yr} \,  (M_{BH}/10^6 M_\odot)^{-1/2} \, (r/ 1\, {\rm pc})^{3/2}$ \citep[see also][]{michikoshi2017}.
In fact, we found that the aggregate disks become gravitationally unstable soon after $S_t$ reaches unity in most cases. 


\
\section{A typical evolution track toward planets}
We investigated the evolution of icy dust particles based on the processes explained in \S 2
to see if the four evolution stages of the dust aggregates are completed, 
by changing the parameters, such as the black hole mass $M_{BH}$, and turbulent 
viscous parameter $\alpha$. 
{The circumnuclear cold gas disk embedded in the geometrically thick torus (see Fig. 1) is assumed to be gravitationally stable. 
We assign a constant Toomre's Q-value, i.e. $Q_g = 2$ with the gas temperature $T_g = 100 $ K in the disk\footnote{Although $Q_g < 1$ is the necessary condition for
the gravitational instability for the $m = 0$ mode perturbation in a thin, uniform disk, the non-axisymmetric modes can be unstable for $Q_g \lesssim 1.5$ \citep[e.g.][]{laughlin1994}. 
Therefore it is safe to assume $Q_g \sim 2$ for a gravitationally stable disk \citep[see also][where the effective Q-value is $\gtrsim 2$ in the multi-phase, quasi-stable gas disk]{wada1999, wada2002}. }.
The hydrogen column density of the gas disk is therefore $N_{\rm H} \simeq 6.2\times 10^{23} \, {\rm cm}^{-2} \, (M_{BH}/10^6 M_\odot)^{1/2} (r/ 1 \, {\rm pc})^{-3/2}  (T_g/100 \, {\rm K})^{1/2} $. The gas mass between $r = 0.1$ pc and 10 pc in the thin disk is
$M_g \simeq 5.7\times 10^4  \, M_\odot \,  (M_{BH}/10^6 M_\odot)^{1/2} (T_g/100 \, {\rm K})^{1/2}$.  The total gas mass in the whole 
torus system of tens parsecs could be comparable to $M_{BH}$, as suggested by recent ALMA observations \citep{izumi2018, combes2019}. }.



\begin{figure}[ht]
\begin{center}
\includegraphics[width = 17cm]{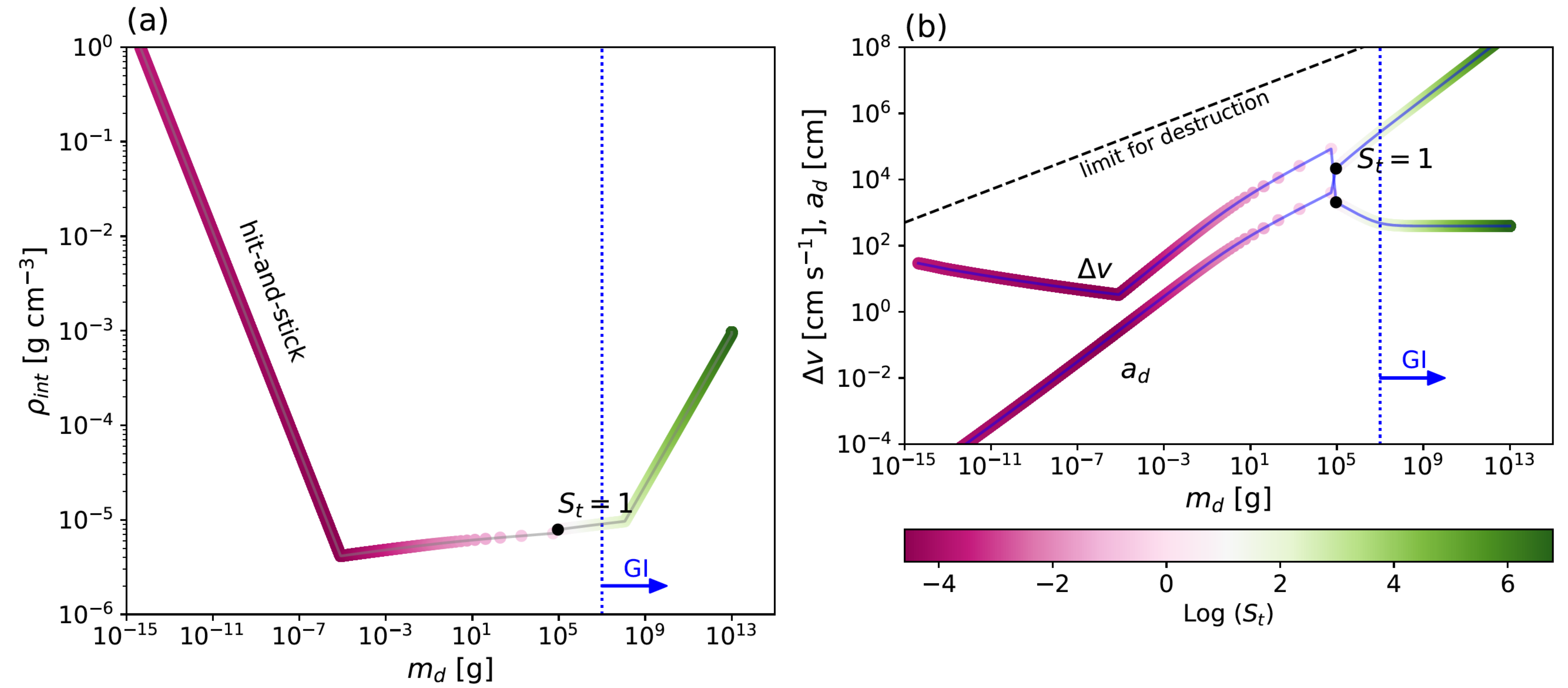} 
\caption{(a) Evolution of the internal density of a dust aggregate $\rho_{int}$  as a function of the aggregate mass $m_d$ for $M_{BH} = 10^7 M_\odot$, $\alpha = 0.1$ and the Eddington ratio is 0.01 (the bolometric luminosity of the AGN is 
$10^{43}$ erg s$^{-1}$),
temperature of the gas $T_g = 100$ K. The dust sublimation radius is located at $r_{sub} = 0.3$ pc and the snow-line is $r_{snow} = 4.7$ pc.  
This plot for the dust at $r = 5.5$ pc.
The aggregates grow by the hit-and-stick process, where the internal density of the aggregates monotonically decreases, that means the aggregates are porous in this phase.  
After the collisional energy exceeds critical energy, the aggregates start to get compressed ($m_d > 10^{-5}$ g).
The color bar represents the Stokes number. For $S_t \sim 1$ the dust aggregates are decoupled from the gaseous turbulence. \\
(b) Collision velocity of the aggregates $\Delta v$ and size of the aggregate $a_d$ as a function of $m_d$. The dashed line shows the limit for the collisional destruction of the aggregates estimated from numerical experiments \citep{wada-k2009}.
After $S_t = 1$ is attained, $\Delta v$ drops and the disk of the aggregates becomes gravitationally unstable to form more massive ``planets", shown by 
the vertical blue dotted line with ``GI" (gravitational instability).
}
\end{center}
\label{fig:rho-mass}
\end{figure}

Figure 2a shows a typical evolution of  a dust aggregate  at $r = 5.5$ pc, just outside of the snow line ($r_{snow} =$ 4.7 pc) around 
a low luminosity AGN with $M_{BH} = 10^7 M_\odot$ and 
the Eddington ratio, i.e. the luminosity ratio to the bolometric luminosity,
$\gamma_{Edd} = 0.01$.
The internal density of the aggregate $\rho_{int}$ is 
plotted as a function of its mass $m_d$. Here, we assume $\alpha = 0.1$.
The internal density decreases monotonically from the monomer's density, $\rho_0 = 1$ g cm$^{-3}$
to $4 \times 10^{-6}$ g cm$^{-3}$, as its mass increases from $m_d \sim 10^{-15}$ g to $ \sim 10^{-5}$ g. 
At that instant, the size of the aggregate becomes $\sim 0.1$ cm.  After this hit-and-stick phase, the fluffy dust aggregates 
keep growing by collisions
in the turbulent gas motion until  $S_t \simeq 1$.
During this phase, the aggregates are compressed by the
collisions, and therefore $\rho_{int}$ gradually increase during this phase ($m_d = 10^{-6}$ g to $10^{5}$ g).
At the end of this phase, 
their size become $a_d \sim 1$ km.
After this stage, the aggregates are compressed by their self-gravity, therefore $\rho_{int}$ increases
quite rapidly as seen below.  

Figure 2b plots collisional velocity $\Delta v$ of the aggregate [cm s$^{-1}$] and its size $a_d$  [cm] as a function of $m_d$. $a_d$ 
monotonically increases for $S_t < 1$. Initially $\Delta v$ is 
50 cm s$^{-1}$ and it decreases 
in the hit-and-stick phase, after which it increases from 10 cm s$^{-1}$ to $\sim 500$ m s$^{-1}$
around $S_t = 1$ in the collisional compression phase. It is far below the limit of the
collisional destruction velocity of aggregates extrapolated
from the numerical experiments of collisions between porous aggregates \citep{wada-k2008a}, which scales with the mass as
$m_d^{1/4}$.

    \begin{figure}[h]
\begin{center}
\includegraphics[width = 11cm]{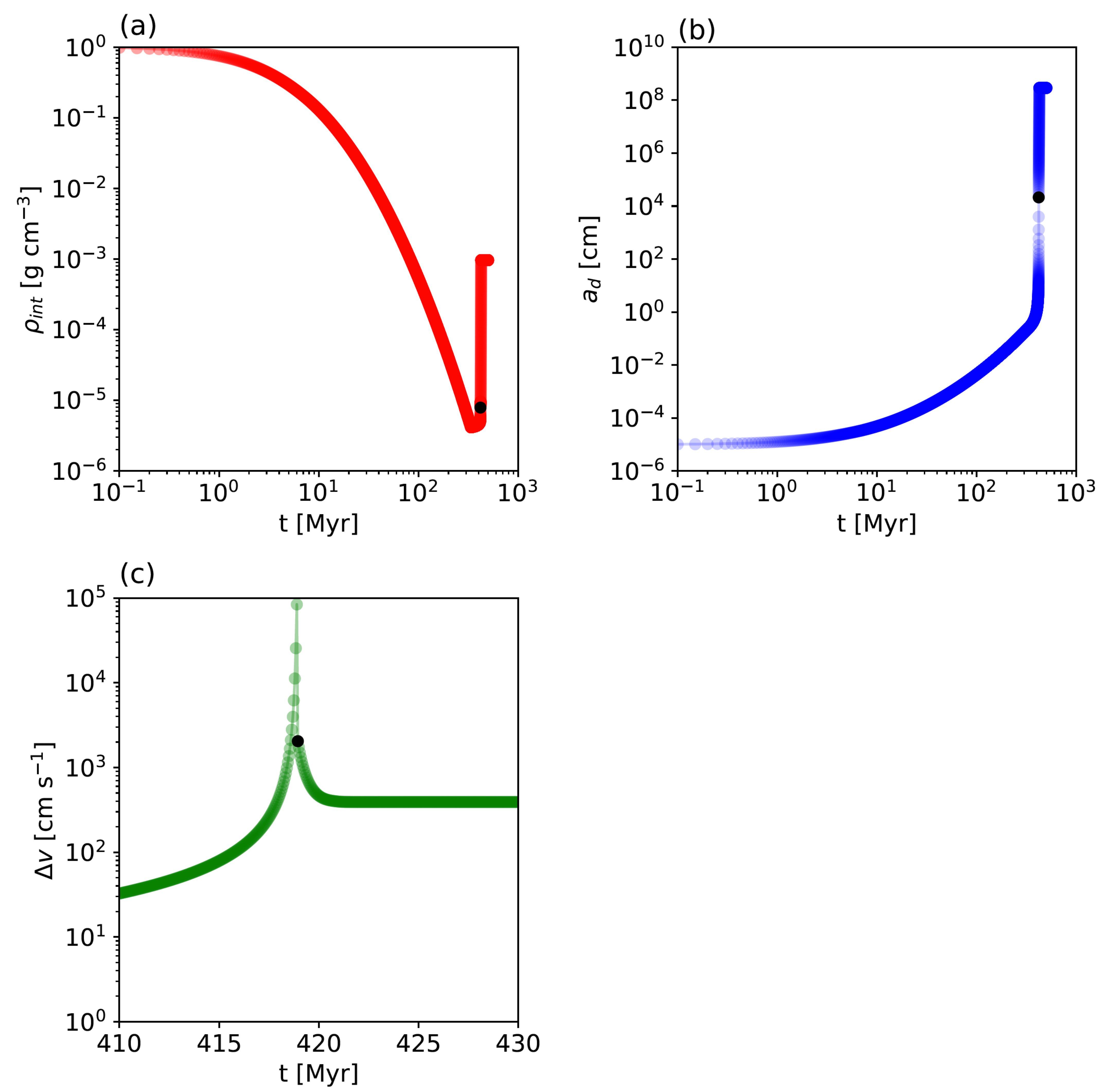} 
\caption{Time evolution of the internal density of the aggregate $\rho_{int}$, size $a_d$
and relative velocity $\Delta v$ for the same model shown in Fig. 2.  The position where $S_t = 1$ is shown by filled circles. }
\label{fig:3}
\end{center}

\end{figure}

Figure \ref{fig:3} shows time evolution of $\rho_{int}$, $a_d$ and $\Delta v$ for the same model
in Fig. 2.  The hit-and-stick phase and the collisional compression phase 
take $\sim 3.8\times 10^8$ yr. 
Soon after the aggregates are decoupled from gaseous turbulence, 
where $S_t \simeq 1$, 
their evolution is determined by
various heating and dissipation (i.e., cooling) processes in the $N$-body system of the aggregates \citep{michikoshi2017}.
Note that Fig. 3b depicts $a_d$ grows up to $\sim 1000$ km, but this does not happen because of the gravitational instability after 
$S_t = 1$. 
Fig. 3c shows that the collisional velocity $\Delta v$ increases rapidly around $S_t \sim 1$, resulting in the rapid growth of the aggregates in the 
mass and density. 
The collisional velocity then dramatically 
decreases to $\sim 4$ m s$^{-1}$ due to collisional loss of their kinematic energy. 
This reduces the Toomre's $Q$-value of the dust disk, and as a result $Q_d < 2$ is attained, therefore the
system of km size aggregates becomes gravitationally unstable (denoted by the dotted lines with ``GI" in Figs. 2a and 2b). 
This leads formation of spiral instabilities and their fragmentation 
followed by collapsing massive ``planets'' consisted of dust. 
This final unstable phase occurs very quickly with a few rotational period (i.e. 4$\times 10^5$ yr at $r = 5.5$ pc for $M_{BH} = 10^7 M_\odot$).  
The mass of ``planets" then would be $m_{pl} \sim \lambda_{GI}^2 \Sigma_d \sim 10 M_E$,
where $\lambda_{GI}$ is the most unstable wavelength for the gravitational instability, and $M_E$ is the Earth mass.
In this model, the total number of the ``planets" outside the snow line ($r =$ 4.7 pc)
to $r =$ 7 pc  is  about $8.5\times 10^4$ and its number density is
$\Sigma_{planet} \sim 10^3 $ pc$^{-2}$.

%
\section{Discussion}
%

We then explored the evolution of the aggregates by changing $\alpha$ and $M_{BH}$.
In Fig. 4, we plot time for which $S_t$ becomes unity as a function of 
$\alpha$ and $M_{BH}$.  
The behavior of the dust growth is basically the same as the model 
with $M_{BH} = 10^7 M_\odot$ and $\alpha = 0.1$ (Figs. 2 and 3),  but 
the time scale to reach the state with $S_t = 1$ depends on $\alpha$ and $M_{BH}$.
For example, for $M_{BH} = 10^6 M_\odot$ and $\gamma_{Edd} = 0.01$,
the snow line is located $r = 1.4$ pc. At $r = 2$ pc, it takes $2.6 \times 10^8$ yr when $S_t$ exceeds unity. 
As shown in eq. (\ref{eq: 7}), the growth time scale is proportional to $\alpha^{-1/2} H_g$, and $H_g \propto M_{BH}^{-1/2} r^{3/2}$. 
The snow line scales as $r \propto L_{AGN}^{1/2} \propto M_{BH}^{1/2}$. Therefore, the growth time scale depends on
$\alpha^{-1/2}  M_{BH}^{1/4}$.
As shown in Fig. 4,  
the time scale becomes $\sim 10^8$ yr for 
 $\log \alpha > {-0.3}$ and $M_{BH} = 10^6 M_\odot$ or  $\log \alpha > {-0.2}$ and $M_{BH} = 10^7 M_\odot$.
In other words, formation of ``planets" can be expected around the circumnuclear 
gas disks in {\it low luminosity Seyfert-type AGNs} rather than quasar-type high luminosity ones with massive SMBHs. 

{The growth time of an aggregate is proportional to $\alpha^{-1/2}$. Therefore, if $\alpha$ in the 
circumnuclear disk is as small as in the protoplanetary disks, i.e. $\alpha \sim 10^{-3} - 10^{-4} $, 
the growth time would be $6\times 10^8 - 2 \times 10^9$ yr, 
which is still smaller than the cosmological time. However, because $\alpha$ is proportional to 
the mass accretion rate (eq.(\ref{eq: alpha})),  the Eddington rate of the central source as a result of the accretion 
would be $\gamma_{Edd} \sim 10^{-4}- 10^{-5}$, which corresponds to very low luminosity AGNs \citep[e.g.][]{ricci2017}.
These imply that the planets could be also formed
within a few giga-years around very faint AGNs.  }

\begin{figure}[h]
\begin{center}
\includegraphics[width = 9cm]{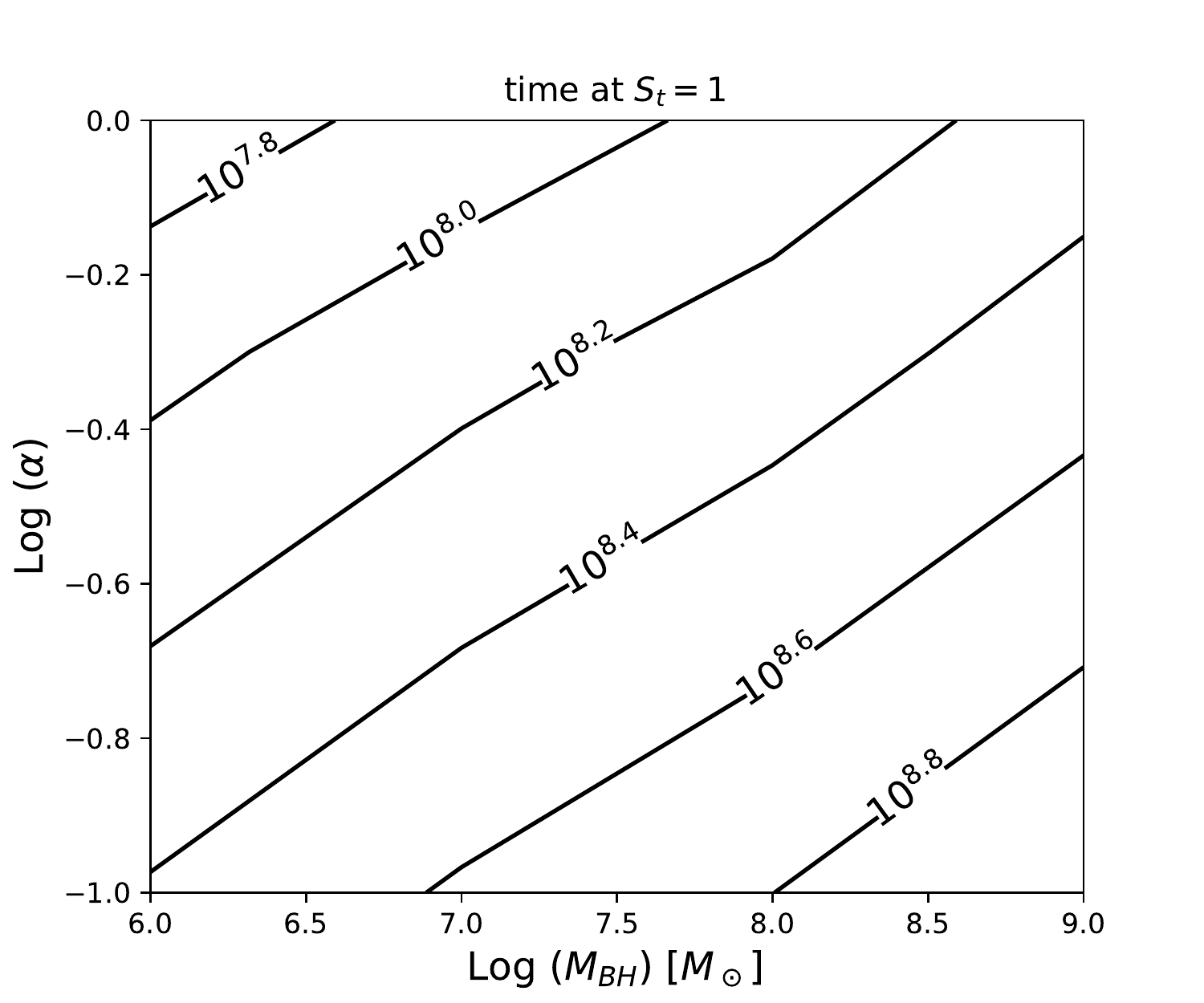} 
\caption{ Evolution time [yr] at  the Stokes parameter $S_t$ becomes unity  as a function of $M_{BH}$ and $\alpha$-parameter \citep{shakura1973}.  After $S_t \sim 1$ is attained, 
the km-sized aggregate system is decoupled from the gaseous turbulence, and it becomes gravitationally unstable, leading to the formation of ``planets" within $\sim 10^6$ yr in this parameter range. }
\label{fig:4}
\end{center}

\end{figure}

{The dust-to-gas mass ratio is assumed to be a standard value, i.e., 0.01 \citep{draine2011}, however
this could be larger by few factors in the high metallicity environment (e.g. $\sim 4 Z_\odot$) around AGNs \citep{groves2006, remy2014}.
In such case, the time scale of the dust evolution can be smaller by few factors than
shown in Fig. \ref{fig:4}.}

Observing planets around SMBHs should be challenging.  The standard techniques to detect exoplanets around stars, i.e.,
Doppler spectroscopy, transit photometry, gravitational micro-lensing, or direct imaging are
hopeless. Photometry by a hard X-ray interferometer in space might be a possible solution, but
the occultation of the accretion disk by the ``planets"  would be hard to distinguish from
the intrinsic time variability of AGNs.
The other, indirect way is detecting spectral changes in the $mm$-wave length due to opacity variation associated with the dust growth as used in protoplanetary disk. The opacity
is roughly proportional to $\rho_{int} \, a_d$, and this increases by more than two orders of magnitude
around $S_t \sim 1$.

\acknowledgments
We would like to appreciate the anonymous referee's valuable comments.
This work was supported by JSPS KAKENHI Grant Number  18K18774.  The authors thank Akio Inoue and Tohru Nagao for suggestions on metallicity and 
the dust-to-gas ratio in AGNs.






\section*{Appendix:  Evolution of dust aggregated in each stage}

Here, we describe detail models of the dust evolution used in \S 3 and \S 4, based on the assumption
 that the elementary processes during the growth of the icy dust monomers to the planetesimals around stars 
 can be applied to the dust particles in the circumnuclear disk around SMBHs. 
 However, there are significant differences between 
the two systems (see Table 1), which could result in very different ``planet" systems around SMBHs.
The evolution of dust particles is divided into four phases as described below.
\subsection*{1. Hit-and-Stick phase}
If the dust aggregates grow by ballistic cluster-cluster aggregation (BCCA), 
the internal structure of the aggregate should be porous (i.e. $\rho_{int} \ll \rho_0 \sim 1$ g cm$^{-3}$), and 
its fractal dimension is $D \simeq 1.9$ \citep{mukai1992, okuzumi2009}.
In this case, the internal density of the aggregates in the hit-and-stick phase \citep{okuzumi2012, kataoka2013} evolves as
\begin{eqnarray}
\rho_{int} = (m_d / m_0)^ {1-3/D}  \rho_0 ,
\end{eqnarray}
where  $m_d$ is the mass of the aggregate, and $m_0$ and $\rho_0$ are the monomer's mass and density, respectively.
We assume that $m_0 =10^{-15}$ g and $\rho_0 = 1$ g cm$^{-3}$. 
In contrast to the protoplanetary disk, the radial motions of the gas and dust are small. For example, the radial velocity of the 
gas disk for the Eddington ratio $\gamma_{Edd} = 0.1$ and the black hole mass $M_{BH} = 10^7 M_\odot$ is $10^{-4} v_K - 10^{-3} v_K$, where $v_K$ is 
the Keplerian rotational velocity (see also eq.(\ref{eq-18}), (\ref{eq-19})).
Therefore, as the first approximation, we assume that the gas and dust surface density distribution ($\Sigma_d (r) = f_{dg} \Sigma_g(r)$) 
does not change during the evolution.

The growth rate for $m_d$ is then
\begin{eqnarray}
\frac{d m_d }{d t} = \frac{2 \sqrt{2 \pi} \, \Sigma_d  \, a_d^2 \, \Delta v}{H_d}, 
\end{eqnarray}
where $a_d$ is the size of the dust aggregate, $\Delta v$ is collisional velocity between the aggregates and $H_d$ is the scale height of the dust disk given as 
\citep{youdin2007, tsukamoto2017}.

\begin{eqnarray}
H_d = \left( 1 + \frac{S_t}{\alpha} \frac{1+ 2 S_t}{1+ S_t} \right)^{-1/2}  H_g,
\end{eqnarray}
where $H_g = c_s/\Omega_K$ is the scale height of the gas disk.

The relative velocity between aggregates $\Delta v$ for $S_t < 1$ can be divided into  two regimes \citep{ormel2007}:
regime I) $ t_s \ll t_{\eta} = t_L \, Re^{-1/2}$, and  regime II) $t_\eta \ll t_s \ll \Omega^{-1}$.
The Reynolds number, 
$R_e \equiv \alpha c_s^2/(\nu_{mol} \Omega)$ with the molecular viscosity $\nu_{mol} \sim \frac{1}{2} c_s \lambda_g$ is
\begin{eqnarray}
R_e &\approx& 3 \times 10^4 \, M_{BH,6}^{-1/2} \, r_1^{3/2} \, c_{s, 1}^{-1} \, Q_g\,  \gamma_{Edd, 0.01},
\end{eqnarray}
where $Q_g$ is the Toomre's $Q$-value for the gas disk.
The eddy turn over time $t_L$ is
$t_L \sim \Omega_K^{-1}$, and $t_\eta \sim t_L$ for the smallest eddy.
For the hit-and-stick phase,  $S_t  \ll  R_e$, then $\Delta v$ obeys the regime I, and 
\begin{eqnarray}
\Delta v_I \sim \sqrt{\alpha} c_s R_e^{1/4} |S_{t,1} - S_{t,2}|
\sim 
\frac{1}{2} \sqrt{\alpha} c_s R_e^{1/4} S_{t} \\
\sim  0.5  \, [{\rm m/s}]   \left(\frac{S_t}{1.5\times 10^{-4}} \right)^{1/2} \, \left( \frac{\alpha}{0.3} \right)^{1/2}  \left( \frac{R_e}{3\times 10^4} \right) \,   c_{s,1} ,
\end{eqnarray}
where $S_{t,1}$ and $S_{t,2}$ are Stokes numbers of two colliding particles. 
We here assume $S_{t,2} \sim S_{t,1}/2$ (Sato et al. 2016).
For regime II, on the other hand, 
\begin{eqnarray}
\Delta v_{II} &\sim& v_L \sqrt{t_{stop}/t_L} \sim \sqrt{\alpha S_t} \, c_s  \\
& \approx & 6.7 \, [{\rm m/s}]   \left(\frac{S_t}{1.5\times 10^{-4}} \right)^{1/2} \, \left( \frac{\alpha}{0.3} \right)^{1/2}  \,   c_{s,1}  
\label{eq: deltav2}
\end{eqnarray}
where $v_L$ is velocity of the largest eddy.

The size of dust aggregates determines how they interact with the gas (e.g. the Stokes parameter is proportional to
$\rho_{int} a_d$ for the Epstein law). Dynamics of the aggregates is affected by their cross sections, which depend on
the internal inhomogeneous structure.
The radius of BCCA cluster $a_{BCCA}$ for large numbers of monomers $N$ is given as 
 $a_{BCCA} \simeq N^{0.5} a_0$ \citep{mukai1992, wada-k2008a, wada-k2009}, and 
this was also confirmed by $N$-body simulations \citep{suyama2012}.

\subsection*{2. Collisional and gravitational compression phases}

The hit-and-stick phase ends, when the rolling energy $E_{roll}$, which is the energy required to rotate a particle around a connecting point by 90$^\circ$,
is comparable to the impact energy, $E_{imp}$ between the porous aggregates.  Beyond this point,  the aggregates start to get compressed
by mutual collisions. Here, we assume $E_{roll} = 4.37\times 10^{-9}$ erg \citep{suyama2012}. 
When the number of monomers in the aggregates exceeds a critical number $N_{crit} \equiv  \beta \frac{ 8 E_{roll}}{m_0 \Delta v^2}
$  with $\beta = 0.5$ \citep{suyama2012},  they are compressed, and the internal density  no longer decreases during the coagulation process. 

The collisional velocity $\Delta v$ between aggregates is determined by the interaction with the turbulence for $S_t < 1$, 
depending on $S_t$ and  $R_e$  \citep{ormel2007}:
For  $S_t  \le R_e$, 
\begin{equation}
\Delta v =
\frac{1}{2} \sqrt{\alpha} \, c_s \, Re^{1/4}  \, S_t   ,
\end{equation} 
or for  $S_t > R_e$, 
\begin{equation}
\Delta v = 
\sqrt{\alpha S_t} \, c_s, 
\end{equation}
where $R_e \equiv \alpha c_s^2/\nu_{mol} \Omega_K$ with the sound velocity of the
gas disk $c_s$. 

The internal density of the aggregated $\rho_{int, f}$ formed of two equal-mass aggregates, 
with density $\rho_{int}$, is calculated according to  \citet{suyama2012}: 
\begin{eqnarray}
\rho_{int, f}^4 = \left( \rho_{int}^4  +  \rho_0^4 \, \frac{E_{imp}}{0.15 N E_{roll}} \right)^{1/4}.
\end{eqnarray}


As the aggregates become more massive for $S_t < 1$,
they start getting compressed by their self-gravity, and 
the internal density evolves as 
$\rho_{int} \propto (\Delta v)^{3/5}  \, m_d^{-1/5}$ \citep{okuzumi2012}. 
This phase ends when the Stokes parameter becomes unity ($S_t \sim 1$). 
Then the aggregates are decoupled from the turbulent gas, and they evolve as
$N$-body system.
 
 \subsection*{3. Evolution of dust aggregates as a $N$-body system}
 
 When $S_t > 1$, the collisional velocity between the aggregates is determined by 
a balance between heating and cooling processes as the $N$-body particles.
According to \citet{michikoshi2016, michikoshi2017},   we solve the following equation to get equilibrium random velocity of 
the dust aggregates $v_d$, 
\begin{eqnarray}
\frac{d v_d^2}{dt} &=& \left( \frac{d v_d^2}{dt} \right)_{grav} + \left( \frac{d v_d^2}{dt} \right)_{turb, stir} +  \left( \frac{d v_d^2}{dt} \right)_{turb, grav} \nonumber \\ 
&-& \left( \frac{d v_d^2}{dt} \right)_{coll}   -\left( \frac{d v_d^2}{dt} \right)_{drag} = 0 . 
\label{eq:19}
\end{eqnarray}
The first three heating terms are due to the gravitational scattering of the particles, stirring by turbulence, and gravitational scattering
by turbulent fluctuation, respectively.  The two cooling terms in eq.(\ref{eq:19})  represent the collisional damping and the gas drag.  
 
 \subsection*{4. Gravitational instability of the aggregates disk}
We investigate gravitational instability (GI) of the disk consisted of 
the dust aggregate at $S_t > 1$ based on 
 the Toomre's $Q$-value defined as  $Q_d \equiv (v_d/\sqrt{3} )\Omega_K/3.36 G \Sigma_d$.  
For the axi-symmetric mode, $Q_d < 1$ is the necessary condition for the liner GI,  but non-axisymmetric mode can be developed for $Q_d \lesssim 2$, 
and the spiral-like density enhancements are formed followed by fragment of the spirals (Michikoshi \& Kokubo 2017), 
which leads formation of planets.
  The mass of the fragments can be estimated as 
$m_{pl} \simeq \lambda_{GI}^2 \Sigma_d$, where
the critical wavelength for GI $\lambda_{GI} = 4 \pi^2 G \Sigma_d/\Omega_K^2$.
The number of "planets" then can be estimated as $N_{pl} \sim  2\pi r/ \lambda_{GI}$. 
We found that the velocity dispersion of the aggregates drops rapidly due to
the cooling terms in eq.(\ref{eq:19}). As a result, the system becomes gravitationally unstable after $S_t = 1$ in
a rotational period.

\end{document}